\documentclass[useAMS,usenatbib]{mn2e}

\usepackage{graphicx} \usepackage{natbib}

\newcommand{\logg}{\mbox{$\log g$}}

\def\m2s2{\hbox{\,m$^{2}$\,s$^{-2}$}} 
\def\Msun{\hbox{$\mathrm{M}_{\odot}$}} 
\def\Rsun{\hbox{$\mathrm{R}_{\odot}$}} 
\def\Mjup{\hbox{$\mathrm{M}_{\rm Jup}$}}
\def\Rjup{\hbox{$\mathrm{R}_{\rm Jup}$}} 
\def\degr{\hbox{$^\circ$}}

\def\1s{$1\,\sigma$} 
 
\def \t0{T$_0$}

\usepackage{times}

\title[A lower mass for the exoplanet WASP-21b]{A lower mass for the
  exoplanet WASP-21b} \author[S.C.C. Barros et al.]{S. C.  C. Barros
  $^{1}$\thanks{E-mail:s.barros@qub.ac.uk}, D. L. Pollacco $^{1}$,
  N. P. Gibson $^{1,2}$, I. D. Howarth $^{3}$,
  F. P. Keenan $^{1}$, \and E. K. Simpson $^{1}$, I. Skillen$^{4}$,  I. A. Steele$^{5}$\\
  $^{1}$Astrophysics Research Centre, School of Mathematics and
  Physics, Queen's University Belfast, University Road, Belfast, BT7
  1NN, UK\\
  $^{2}$Department of Physics, University of Oxford, Denys Wilkinson
  Building, Keble Road, Oxford
  OX1 3RH, UK \\
  $^{3}$Department Physics and Astronomy, UCL, Gower
  Street, London WC1E 6BT, UK  \\
  $^{4}$Isaac Newton Group of Telescopes, Apartado de Correos 321,
  E-38700 Santa Cruz de la Palma, Tenerife,
  Spain\\
  $^{5}$Astrophysics Research Institute, Liverpool John Moores
  University, CH61 4UA, UK}

\begin{document}

\date{Accepted . Received ; in original form }

\pagerange{\pageref{firstpage}--\pageref{lastpage}} \pubyear{2002}

\maketitle

\label{firstpage}

\begin{abstract}
  We present high precision transit observations of the exoplanet
  WASP-21b, obtained with the RISE instrument mounted on 2.0m
  Liverpool Telescope.  A transit model is fitted, coupled with an
  MCMC routine to derive accurate system parameters. The two new high
  precision transits allow to estimate the stellar density directly
  from the light curve. Our analysis suggests that WASP-21 is evolving
  off the main sequence which led to a previous overestimation of the
  stellar density.   Using isochrone interpolation, we find a stellar
  mass of $0.86 \pm 0.04 \Msun$ which is significantly lower than
  previously reported ($1.01 \pm 0.03 \Msun$ ). Consequently, we find
  a lower planetary mass of $0.27 \pm 0.01 \Mjup$.  A lower
  inclination ($87.4 \pm 0.3$ degrees) is also found for the system
  than previously reported, resulting in a slightly larger stellar
  ($R_* =1.10 \pm 0.03\,$\Rsun) and planetary radius ($R_p = 1.14 \pm
  0.04\, $\Rjup).  The planet radius suggests a hydrogen/helium
  composition with no core which strengthens the correlation between
  planetary density and host star metallicity. A new ephemeris is
  determined for the system, i.e., \t0 $=2455084.51974 \pm 0.00020$
  (HJD) and $P=4.3225060 \pm 0.0000031\,$ days. We found no transit
  timing variations in WASP-21b.

\end{abstract}


\section{Introduction}

Transiting planet systems are valuable because their geometry enables
us to estimate accurate planetary properties. Time-series photometry
during the transit allows us to derive the orbital inclination and the
relative radii of the host star and planet. Combining this with radial
velocity variations and stellar parameters, allows us to derive the
absolute mass of the planet. Hence, the bulk density of the planet can
be estimated with good accuracy, giving us insight into its
composition \citep{Guillot2005, Fortney2007}, thus placing constraints
on planetary structure and formation models. Given the remarkable
diversity in the structure of large planets, it is important to obtain
planetary parameters which are as accurate as possible.  However,
obtaining high signal-to-noise transit observations is difficult and
consequently even some of the brightest stars with planets are lacking
good quality light curves and, hence, have poorly determined planetary
parameters.

The RISE (Rapid Imager to Search for Exoplanets) instrument, mounted
on the 2.0m Liverpool telescope \citep{rise2008, Gibson2008} was
designed for exoplanet transit observations. Its main scientific
driver was the detection of transit-timing variations and hence the
search for low-mass companions to ``hot Jupiters''. RISE has a rapid
readout frame transfer CCD and in 2$\times$2 binned mode has a readout
time of less than 1 s. This implies that for exposures longer than 1
s, dead time is negligible substantially increasing the time on
target. However, most exoplanet host stars are relatively bright and
saturate the CCD for 1 second exposure. To avoid dead time losses,
RISE observations are always defocussed
(e.g. \citealt{Gibson2008,Joshi2009}).  Defocussed photometry
observations have also the advantage of spreading the PSF over a
larger number of pixels, thereby decreasing flat-fielding errors. RISE
is therefore ideal for obtaining high quality transit light curves for
exoplanets.

WASP-21b is a Saturn-mass planet with $M_p = 0.30 \pm 0.01 $ \Mjup\ in
a 4.3 day circular orbit \citep{Bouchy2010}. Its host star is a G3V
type with $M_*=1.01 \pm 0.03\,$\Msun\ , $T_{eff} = 5800 \pm 100\,$K
and a low metallicity, {[M/H]} =$-$0.4 $\pm$ 0.1. It was discovered by
the SuperWASP-North survey \citep{Pollacco2006} in its 2008-2009
observing campaign. \citet{Bouchy2010} argue that WASP-21 is a member
of the galactic thick disc because of its low metal abundances,
velocity relative to the Sun and age $ \sim 12$Gyr, which are similar
to the thick disc population. WASP-21b is among the lowest density
planets, $ \rho_p=0.24\pm 0.05 \rho_J$ \citep{Bouchy2010}, and has one
of the lowest metallicity host stars. Therefore, its properties are
particularly important for irradiation models.  The current parameters
of the system \citep{Bouchy2010} are based on the SuperWASP discovery
photometry and a partial transit light curve taken with RISE. However,
the lack of a high precision complete transit light curve required the
assumption of the main sequence mass-radius relation which tends to
bias the estimate of the inclination. Furthermore, the age derived for
WASP-21 is longer than the main-sequence life time of a $1.01 \Msun$
star. This suggests that WASP-21 could be evolved which would
invalidate the main-sequence assumption and bias the parameters of the
system. To test the main-sequence assumption we obtained further
observations of WASP-21.

In this paper, we present transit observations of WASP-21b with RISE
including a full transit light curve. Our high precision light curves
allow us derive the planetary and stellar radii without assuming the
main-sequence mass-radius relation for the host star. We describe our
observations in Section 2. In Section 3, we discuss our transit model
and present the updated parameters of the system in Section
4. Finally, we discuss and summarise our results in Section 5.

\section{Observations}

WASP-21b was observed with RISE \citep{rise2008} mounted at the
auxiliary Cassegrain focus of the robotic 2.0m Liverpool Telescope on
La Palma, Canary Islands. This is a focal reducer system utilizing a
frame transfer e2v CCD sensor. The detector has a pixel scale of 0.54
arcsec/pixel that results in a 9.4 $\times$ 9.4 arcmin field of
view. RISE has a wideband filter covering $\sim 500$--$700\,$nm which
corresponds approximately to V+R. The instrument has no moving parts.

The Liverpool Telescope has a library of flat fields which are taken
manually every couple of months.  RISE flats are taken during twilight
at different rotator angles so that there is a uniform illumination of
the CCD. The exposure times of the images are automatically adjusted
so that the peak counts in the individual flats are below the
non-linearity limit of the CCD at 45000 counts. Typically, the
individual flats have between 20000 and 40000 counts. Due to the fast
readout, we can obtain approximately 200 flat frames in a run, these
are combined to create a master flat. For each observation run we use
the master flat that is closest in time, although we note that these
are very stable.

On 2009-09-09 we obtained a full transit of WASP-21b. A total of 6581
exposures in the $2 \times 2$ binning mode with an exposure time of
2.7 seconds were taken. The telescope was defocussed by -1.2mm which
resulted in a FWHM of $\sim 11$ \arcsec. For defocussed photometry,
the star profiles are not Gaussian. However, we found that, in our
case, a Gaussian provided a good fit to the wings of the star profile,
and could be use as a rough estimate of the profile width.  Therefore,
we estimated the FHWM in the usual way by cross-correlating a Gaussian
profile with that of the star.

A second full transit observation of WASP-21b was attempted on
2010-11-24. In this case, deteriorating weather terminated the
observations shortly after the mid-transit, by which time, 4008
integrations had been obtained. During these observations, the FWHM
was $\sim 12.5$ \arcsec.

Both data-sets were reduced using the ULTRACAM pipeline
\citep{Ultracam} which is optimized for time-series photometry.
Initially, we bias subtracted the data while we investigated
systematic effects that were introduced by the flat fielding process.
We performed differential photometry relative to five comparison stars
in the field, confirmed to be non variable, and we sampled different
aperture radii and chose the aperture radius that minimised the
noise. For the first night, we used a 22 pixel aperture radius ($\sim
12$\arcsec), and for the second transit, a 32 pixel aperture radius
($\sim 17$\arcsec). The photometric errors include the shot noise,
readout and background noises.

We also included in our analysis, the previously published egress of
WASP-21 taken with RISE \citep{Bouchy2010}. For consistency, we
re-reduced the original data using the same method as for the other
two observations.  On 2008-10-07, 2220 exposures of 5 sec duration
were taken. We estimated a FWHM of $\sim 2.7$ \arcsec, therefore, the
level of defocussing was lower than in our observations. The best
aperture radius was found to be 15 pixel ($\sim 8$\arcsec). Our
results agree well with the previous published light curve.

\begin{figure}
  \centering
  \includegraphics[width=\columnwidth]{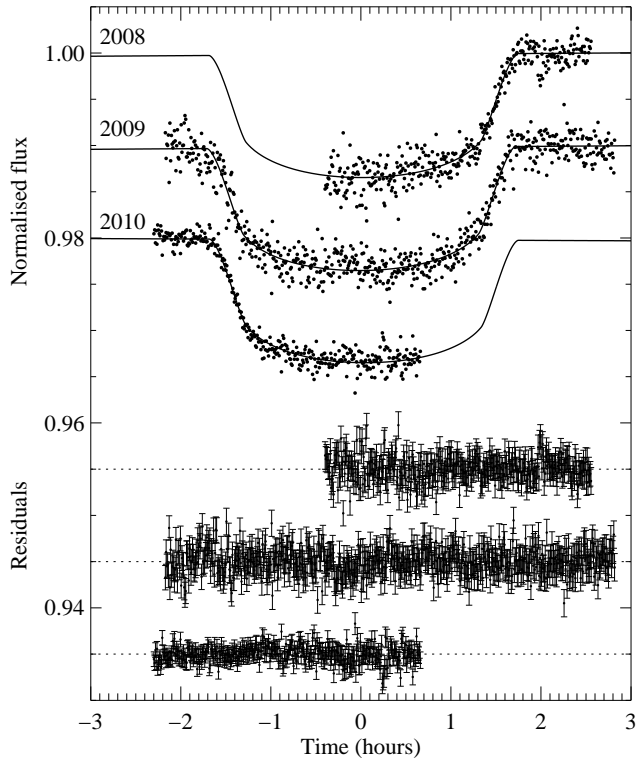}
  \caption{Phase-folded RISE light curves for WASP-21. From top to
    bottom in chronological order; 2008 October 07, 2009 September 09
    and 2010 November 24. We superimpose the best-fit transit model
    and also show the residuals for each light curve at the bottom of
    the figure. The data are binned into 30 second periods, and bins
    displaced vertically for clarity. The individual RISE light curves
    plotted here are available in electronic form at CDS.}
  \label{photolc}
\end{figure}

The final high precision photometric light curves are shown in
Figure~\ref{photolc} along with the best-fit model described in
Section~\ref{model}. We overplot the model residuals and the estimated
uncertainties which are discussed in Section~\ref{errors}.

\subsection{Optimum exposure time for RISE}
As mentioned above, defocusing is commonly used in exoplanet transit
observations.  \citet{Southworth2009} calculated the optimum exposure
time for the DFOSC imager mounted on the 1.54m Danish Telescope. We
follow the same procedure and apply it to RISE mounted on the
Liverpool Telescope and hence, we account for readout noise, photon,
background and scintillation noise. Similar to \citet{Southworth2009},
we do not include flat-fielding noise, assuming that the profile
position is stable.

The key difference is that RISE is a frame transfer CCD whose dead
time is the frame transfer time, 35 milliseconds for observations
longer than 1 second. For the brightest comparison star in our field,
($V \approx 9$), we found the optimum exposure times with RISE are
approximately 2.7, 7.8, 10.8 seconds during bright, gray and dark
time, respectively.

We iterate that the improvement in signal-to-noise for defocussed
observations, reported by \citet{Southworth2009} is only due to
deadtime losses; hence, the defocussing needed is proportional to the
CCD readout time. If the deadtime was zero the best theoretical
signal-to-noise would always be for focused observations, mainly due
to the increase in background noise for wider profiles.

Moreover, in our case, the improvement on signal-to-noise between 1
second and 10.8 seconds exposure times is quite small on the order of
$10\,$ppm per $30\,$sec bin.  As we will see below, the strongest
reason for defocussing is to minimise systematic noise which, due to
its nature, is not accounted for in the calculation and can
substantially increase the noise in a transit light
curve. Figure~\ref{complc} shows systematic noise variations larger
than 400 ppm.

\section{Data analysis}

\subsection{Systematic noise}

\begin{figure}
  \centering
  \includegraphics[width=\columnwidth]{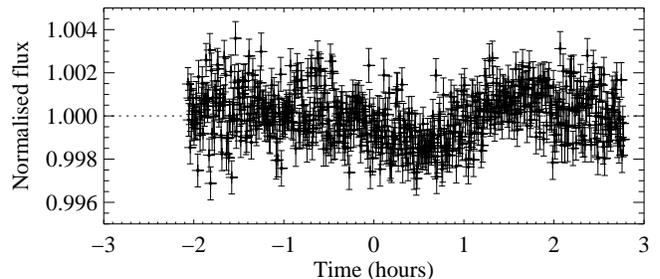}
  \caption{Light curve of the brightest comparison star for the 2009
    September 09 observation relative to the ensemble of comparison
    stars used in WASP21b final light curve. It shows systematic noise
    with an amplitude of 400 ppm. This comparison was not used in the
    final light curve of WASP-21. We also overplot the photometric
    errors.}
  \label{complc}
\end{figure}

Exoplanet transit observations are often dominated by systematic
noise. Therefore, to improve the precision of the light curves it is
important to determine and minimise this noise source. For the 2009
September 09 observations, the brightest comparison star (c1) on the
field was affected by systematic noise.  This can clearly be seen in
Figure~\ref{complc}, where we show the flux of c1 relative to the
ensemble of comparison stars used in the final 2009 WASP-21 light
curve. This shows a variation of 400 ppm.  We found that this
systematic noise was correlated with the star position in the CCD,
which during the transit observation varied by 10 pixels in the $x$
direction and 8 in the $y$. Given that we used an aperture radius of
22 pixels, this implies that only half of the pixels used to perform
aperture photometry were common for the duration of the
observation. Hence, we concluded that the systematic noise was due to
variations in the pixel-to-pixel sensitivity which were not corrected
by flat fielding. In fact, the systematic noise is slightly higher if
we flat field the data. Our master flat is a combination of 150
frames, each with a mean of 35000 counts. The uncertainty in this flat
is 0.5 millimags per pixel which is smaller than the photometric error
($\sim 4.4$ milimags per unbinned point) and the observed systematic
noise. After careful analysis of the data, we found that the c1
comparison star crossed a reflection feature in the CCD that is
rotator dependent (LT is on an alt-azimuth mount) and thus was not
corrected by flat fielding. This experience demonstrates the
importance of good guiding in decreasing the sources of systematic
noise. If the observations were performed in focus and assuming the
seeing was 1 arcsec, the FHWM would have been $\sim 2$ pixels. Using
an aperture radius of $1.5 \times\,$ FWHM $=3\,$ pixels, it would have
implied that there were no common pixels during the
observations. Therefore, we infer, if the observations were focused,
the amount of systematic noise would have doubled. Note that the defocussing does not affect the guiding since the guide camera is always kept in focus.

After this incident the RISE instrument was upgraded. The source of
the reflected feature was identified and removed from the instrument
field of view. We also improved the telescope guiding system's
stability. This led to an improvement in the precision of the light
curves which is evident in the latest light curve of WASP-21 taken
after the upgrades (see Fig. 1). In the November 2010 observations the
variation in position is less than 2 pixels in the $x$ direction and 4
pixels in the $y$.

\subsection{Photometric errors}
\label{errors}
An accurate estimate of the photometric errors is important to obtain
reliable system parameters.  Our first estimate of errors for each
light curve includes only the shot noise, readout and background
noise, which underestimates the true errors. To obtain a more reliable
estimate we begin by scaling the errors of each light curve so that
the reduced $\chi^2$ of the best fitting model is 1.0. This resulted
in the multiplication of the errors by $1.97$, $1.22$ and $1.44$, for
the 2008, 2009 and 2010 light curves, respectively. We then calculated
the time-correlated noise following the procedure from
\citet{gillon2009}.  Using the residuals of the best fit model, we
estimated the amplitude of the red noise, $\sigma_r$ to be $150\,$ppm,
$250\,$ppm and $150\,$ppm, for the 2008, 2009 and 2010 light curves,
respectively. These were added in quadrature to the rescaled
photometric errors and were used in the final Markov Chain Monte Carlo
(MCMC) chains. However, \citet{carter2009} found that this ``time-averaging'' method of estimating the correlated noise can still underestimate the uncertainties by 15-30 per cent.

\subsection{Determination of system parameters}
\label{model}
To determine the planetary and orbital parameters, we fitted the three
RISE light curves of WASP-21b simultaneously. We used the
\citet{Mandel2002} transit model parametrised by the normalised
separation of the planet, $a/R_*$, ratio of planet radius to star
radius, $ R_p/R_* $, orbital inclination, $i$, and the transit epoch,
$T_0$, of each light curve. Our model was originally developed to
measure transit timing variations of exoplanets. Following
\citet{Bouchy2010} that found no evidence for a significant orbital
eccentricity of WASP-21b we adopt a circular orbit.  We included the
quadratic limb darkening (LD) coefficients for the RISE filter V+R
from the models of \citet{Howarth2010}: $a=0.45451$ and
$b=0.210172$. These were calculated for $T_{eff} = 5800\,$K, \logg=4.2
and [M/H] =$-$0.5 to match the stellar parameters from
\citet{Bouchy2010}.  We initially kept the limb darkening parameters
fixed during the fit. For each light curve, we included two extra
parameters to account for a linear normalization. Therefore, 12
parameters were fitted. Besides the linear normalization, no extra
trends were removed from the light curve.

To obtain the best fit parameters and uncertainties, we used a MCMC
algorithm (e.g. \citealt{Tegmark2004,Cameron2007,Gibson2008}). We
begin by calculating the $\chi^2$ statistic of a set of proposed
parameters,
\begin{equation}
  \chi^2=\displaystyle\sum\limits_{j=1}^N {\frac{(f_j-m_j)^2}{\sigma^2_j}},
\end{equation}
where $f_j$ is the flux observed at time $j$, $ m_j$ is the model flux
and $\sigma_j$ is the uncertainty of each $f_j$ as described in
Section~\ref{errors}.  At each step in the MCMC chain, each proposed
parameter is perturbed by a random amount which we call a ``jump
function''. Each jump function is proportional to the uncertainty of
each parameter multiplied by a random Gaussian number with mean zero
and unit standard deviation. The new parameter set is accepted with
probability,
\begin{equation}
  P=min \left( 1, exp\left(\frac{-\Delta\chi^2}{2}\right) \right),
\end{equation}
where $\Delta\chi^2$ is the difference in the $\chi^2$ of subsequent
parameters sets. Note, the new parameter set is always accepted if its
$\chi^2$ is lower than the previous parameter set ($P=1$). The jump
functions are scaled by a common factor in order to ensure that $25\%$
of the steps are accepted, as suggested by \citet{Tegmark2004}. To
estimate the uncertainty of each parameter and calculate the jump
functions, an initial MCMC fit was performed. With these jump
functions, we computed seven MCMC chains each of 150 000 points and
different initial parameters. The initial 20\% of each chain that
corresponded to the burn in phase were discarded and the remaining
parts merged into a master chain. We estimated the best fit parameter
as the mode of its probability distribution and the 1 $\sigma$ limits
as the value at which the integral of the distribution equals 0.341\%
from both sides of the mode. We computed the \citet{Gelman92}
statistic for each fitted parameter and concluded that chain
convergence was good.

To test how the limb darkening coefficients affect the derived system
parameters, we repeated the MCMC procedure also fitting for the linear
LD ``$a$'' which is the most sensitive to the observing filter. The
quadratic LD coefficient, $b=0.210172$, was kept fixed, because as
reported by \citet{Gibson2008}, the high precision of the RISE light
curves is not enough to fully constrain the LD coefficients (i.e., the
MCMC does not converge when fitting both coefficients). We restricted
the linear LD coefficient to be the same for all the light curves
since they were all taken with the same filter. Therefore, in the
second MCMC procedure we fitted 13 parameters. We estimated $a = 0.337
\pm 0.034$.

\section{Results}

Comparing the fitted and fixed LD solutions, we concluded that
although the fitted linear LD coefficient is statistically
significantly different from the theoretical value, this does not
affect the derived system parameters to any extent. The two solutions
are within $1.5\sigma$ of each other. Contrary to what was found by
other authors (e.g., \citealt{Gibson2008,Southworth2008}), the
uncertainties of the fitted LD solution are slightly smaller than
those of the fixed solution. The $\chi^2$ of the fitted LD solution is
similar to the fixed LD solution which does not justify the addition
of an extra free parameter in the fit. Consequently, we conclude that
our light curve is of insufficient quality to better constrain the
linear LD relative to that achieved by theoretical models and we
choose to present the fixed LD solution.

The estimated transit times, combined the original ephemeris
\citep{Bouchy2010} were used to update the linear ephemeris,
\begin{equation}
  T_{t} (HJD) = T(0) + EP.
\end{equation}
We found $P=4.3225060 \pm 0.0000031$ and $T_0 =2455084.51974 \pm
0.00020$ which was set to the mid transit time of the 2009 light
curve. This ephemeris was used in the final MCMC procedures.

For future reference, the time residuals from the linear ephemeris are
given in Table~\ref{ttv}.  We conclude that the time residuals of
WASP-21b are consistent with a linear ephemeris.

\begin{table}
  \centering 
  \caption{Time residuals from the linear ephemeris.}
  \label{ttv}
  \begin{tabular}{ccc}
    \hline
    \hline
    Epoch & Time residuals (sec) & Uncertainty (sec) \\
    \hline
    -78 &    11 & 40 \\
    0   &    -7 & 24 \\
    102 &   8   & 30 \\
    \hline
  \end{tabular}
\end{table}

The geometric system parameters of WASP-21 and the 1$\sigma$
uncertainties derived from the MCMC analysis with fixed limb darkening
coefficients are given in Table~\ref{mcmc}. These parameters are
directly measured from the transit light curve and are only weakly
dependent of stellar properties through the limb darkening
coefficients. Note, all the derived parameters presented in
Table~\ref{mcmc} were calculated at each point of the chain.
Therefore, the final derived values and errors were determined from
their probability distribution as done for the fitted values. We
obtain a significantly lower density than was previous reported in the
discovery paper $ \rho_* = 0.84 \pm 0.09 \rho_{\odot} $).

\begin{table}
  \centering 
  \caption{WASP-21 system parameters derived from the mcmc}
  \label{mcmc}
  \begin{tabular}{lccl}
    \hline
    \hline
    Parameter & Value   \\
    \hline
    Normalised separation $a/R_*$ & $9.68^{+0.19}_{-0.30}$\\
    Planet/star radius ratio $ R_p/R_* $ & $0.10705^{+0.00082}_{-0.00086}$  \\   
    Orbital inclination $ I $ [degrees] & $  87.34 \pm 0.29$ \\    
    Impact parameter $ b $ [$R_*$]  & $ 0.458^{+0.043} _{-0.036}$  \\      
    Transit duration $T_T$ [days] & $ 0.1430^{+0.0013}_{-0.0010}$\\     
    Stellar density $ \rho_* $ [$\rho_{\odot}$] & $ 0.652^{+0.041}_{-0.060}  $ \\
    \hline
  \end{tabular}
\end{table}

\subsection{Stellar mass and age}
To obtain the stellar and planetary physical properties, the geometric
parameters have to be scaled with the stellar mass.  The new high
quality transit light curves give a direct estimate of the stellar
density. This allows a more accurate estimation of stellar mass than
log g derived from spectral analysis \citep{Sozzetti2007}.  Currently
there are two main methods to derive the stellar mass from the stellar
density.  The first uses isochrones and mass tracks from stellar
models \citep{Sozzetti2007} and the second uses an empirical
calibration derived from stellar eclipsing binaries
\citep{Torres2010,Enoch2010}.

\citet{Bouchy2010} derived the stellar mass through the empirical
calibration between $T_{eff}$, $ \rho_* $ and [Fe/H]
\citep{Torres2010} with the parametrisation of \citet{Enoch2010}.
Following the same procedure, with the improved $\rho_*$, we derive a
stellar mass of $1.02 \pm 0.05 \Msun$.  In Table~\ref{mcmc2} we
present the mass and radius of WASP-21 and WASP-21b and the 1$\sigma$
uncertainties derived from the MCMC for a stellar mass of $1.02 \pm
0.05 \Msun$.  We obtain a significantly larger stellar and planetary
radius than previously reported. This is due to the main-sequence
assumption in the previous analysis which as discussed below is found
to be invalid.

\begin{figure}
  \centering
  \includegraphics[width=\columnwidth]{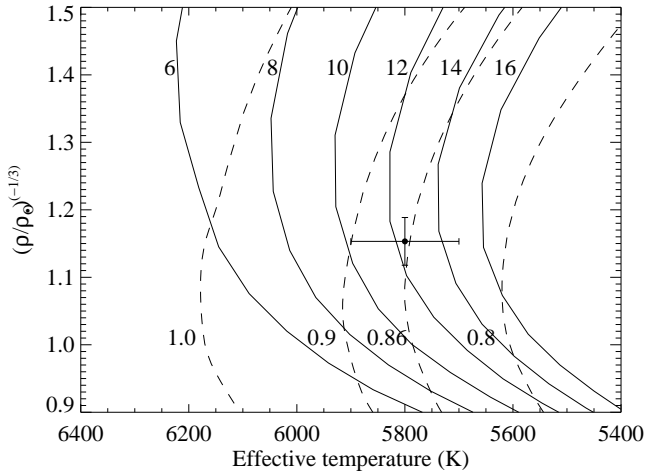}
  \caption{ Isochrone models (solid lines) from \citet{Demarque2004}
    for WASP-21 using {[Fe/H]} =$-$0.47 and {[M/H]} =$-$0.4 from
    \citet{Bouchy2010}.  The age in Gyr is marked in the left of the
    respective model. We also show the mass tracks (dashed lines) for
    a stellar mass of $1.0, 0.9, 0.86$ and $0.8 \Msun$. We overplot
    the $T_{eff} = 5800$ value adapted from \citet{Bouchy2010} and the
    new $( \rho_*/ \rho_{\odot})^{-1/3} $.  }
  \label{isocrones}
\end{figure}

\begin{figure}
  \centering
  \includegraphics[width=\columnwidth]{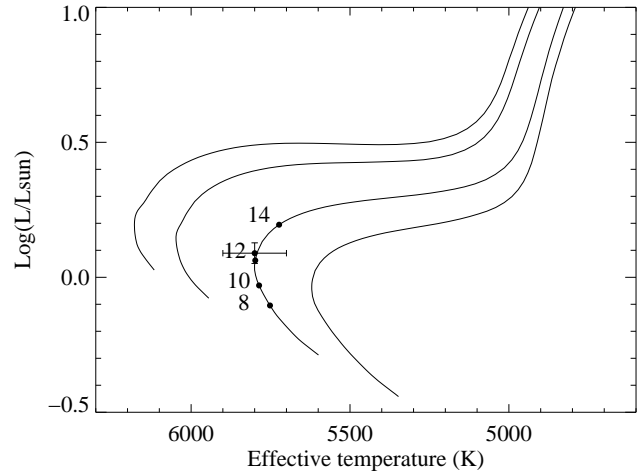}
  \caption{Evolutionary mass tracks from \citet{Demarque2004} for the
    same stellar parameters as Fig.~\ref{isocrones} from left to right
    for stellar masses of 1.0, 0.95, 0.86 and 0.8 \Msun\ . WASP-21
    position is in the 0.86 \Msun\ evolutionary mass track for which
    we also show the 8, 10, 12, and 14 Gyr points.}
  \label{masstrack}
\end{figure}

We also estimate the stellar mass from stellar models by interpolating
the Yonsei-Yale stellar evolution tracks by \citet{Demarque2004} using
the metallicity from \citet{Bouchy2010}. These evolution tracks are
plotted in Figure~\ref{isocrones} along with the position of
WASP-21. From the isochrones, we estimate a lower mass of $0.86 \pm
0.04 \Msun$ and an age of $12 \pm 2\,$Gyr for WASP-21. In Figure
\ref{masstrack}, we also show the evolutionary tracks for stellar
masses of 1.0, 0.95, 0.86 and 0.8 \Msun\ adapted from
\citet{Demarque2004}. These suggest that WASP-21 is close to, or is
already in the hydrogen-shell burning phase and hence is evolving off
the main-sequence. This implies that the assumption of a main sequence
mass-radius relationship in the original analysis of
\citet{Bouchy2010} is faulty.

There is a significant difference between the mass derived from
evolutionary models, $M_*=0.86 \pm 0.04 \Msun$, and the mass derived
from the empirical calibration, $M_* = 1.02 \pm 0.05 \Msun$. In the
past, the \citet{Torres2010} calibration was found to be in agreement
and a more straight-forward alternative to the stellar models
\citep{Torres2010,Enoch2010}. Moreover, it has the advantage that it
can be directly included in a transit fitting procedure
\citep{Enoch2010}. However, recently the same discrepancy between
empirical and isochrone masses was also found for WASP-37
\citep{Simpson2011} and WASP-39 \citep{Faedi2011}. The
\citet{Torres2010} eclipsing binaries sample used for calibrating
their relationship does not contain many low-metallicity systems, in
particular in the low-mass regime. Therefore, this suggests that the
\citet{Torres2010} calibration might not hold for metal poor stars
specially in the low mass regime. For these reasons, for WASP-21, we
favour the lower mass derived from the evolution models. For a stellar
mass of $M_*=0.86 \pm 0.04 \Msun$ we present the stellar and planetary
radii for the WASP-21 system in Table~\ref{mcmc2} along with the
1$\sigma$ uncertainties derived from the MCMC analysis.

To summarise, we derive a lower stellar mass, $0.86 \pm 0.04 \Msun$,
and a lower planetary mass, $0.27 \pm 0.01 \Mjup$.  We estimate the
inclination of the orbit to be $87.3 \pm 0.3$ degrees.  The radius of
the star is found to be $1.10 \pm 0.03\,$ \Rsun\ and the planet radius
is $1.14 \pm 0.04\,$ \Rjup\ , yielding a planetary density of $ 0.18
\pm 0.02\, \rho_J\ $.

\begin{table}
  \centering
  \caption{WASP-21 system stellar and planetary parameters derived using either  the empirical calibration of \citet{Torres2010} or the YY stellar models \citep{Demarque2004}. }
  \label{mcmc2}
  \begin{tabular}{lccccl}
    \hline
    \hline
    &Torres models & YY models   \\
    \hline
    Stellar mass $ M_* $ [\Msun] & $ 1.02 \pm 0.05  $ & $ 0.86 \pm 0.04  $  \\
    Stellar radius $ R_* $  $ R_* $ [\Rsun] &$    1.161^{+0.037}_{-0.024}$ & $    1.097^{+0.035}_{-0.022}$\\
    Stellar surface gravity  $ \log g_* $ [cgs]   & $ 4.32 \pm 0.02 $ & $ 4.29 \pm 0.02 $ \\
    Orbital semimajor axis $ a $ [AU] & $ 0.052 \pm 0.001 $ & $ 0.0494 \pm 0.0009 $ \\ 
    &    &      &  \\
    Planet mass  $ M_p $ [\Mjup]  & $ 0.30  \pm  0.01 $ & $ 0.27  \pm  0.01 $    \\
    Planet radius $ R_p $ [\Rjup]  & $  1.210^{+0.048}_{-0.032}$  & $  1.143^{+0.045}_{-0.030}$  \\
    Planet density  $ \rho_p $ [$\rho_J$] & $ 0.171^{+0.014}_{-0.018}$  & $ 0.181^{+0.015}_{-0.020}$ \\
    Planet surface gravity  $ \log g_P $ [cgs]   & $ 2.71 \pm 0.02 $ & $ 2.71 \pm 0.02 $ \\
    \hline
  \end{tabular} 
\end{table}

\subsection{Eccentricity}

\citet{Bouchy2010} found the eccentricity to be statistically
  indistinguishable from zero, i.e. the $\chi^2$ does not
  significantly improve when adding the two additional parameters to
  the circular model. For these cases, allowing the eccentricity to
  float tends to overestimate the eccentricity \citep{Lucy1971}.
  Hence, \citet{Bouchy2010} adopted a circular orbit. Assuming a tidal
  dissipation parameter between $10^5$ and $10^6$, the circularisation
  timescale for WASP-21b is approximately between $0.017$ and
  $0.17\,$Gyr, respectively. Since this is much shorter than the
  derived age for the system we expect a circular orbit. However, if
  the orbit not circular assuming a zero eccentricity results in
  underestimated uncertainties.  Therefore, it is interesting to
  investigate the effect of a small non-zero eccentricity in the
  system parameters and their uncertainties. As an example, we assume
  an eccentricity of $0.04\pm 0.04$ which is consistent with the
  discovery paper.  We repeated the MCMC procedure allowing the
  eccentricity to float. Because the transit light curves do not
  constrain the eccentricity, we include a prior on the eccentricity
  of the form:
  \begin{equation}
    \frac{(ecc-ecc_{0})^2}{\sigma^2_{ecc}},
  \end{equation}
  where we assume $ecc_{0}=\sigma_{ecc}=0.04$. This prior is added to
  equation 1 at each step of the chain.  From the posterior
  eccentricity distribution we obtain an eccentricity of $0.038\pm
  0.036$ which is close to the input value.

The maximum effect of the eccentricity upon the derived
  parameters corresponds to the case where the transit occurs close to
  periastron ($\omega=90\degr)$ or apastron ($\omega= - 90\degr)$.
  Hence, in order to investigate the maximum deviation from a circular
  orbit we assume $\omega = 90 \degr$. For this particular case by
  assuming a circular orbit we would be overestimating $a/R_*$, $inc$
  and $\rho_*$, and underestimating the stellar and planetary masses
  and radii. The opposite would have happen if we have assumed $\omega
  = -90 \degr$. 
  
The derived eccentric solution is within one sigma of the
  circular solution and the uncertainties of $a/R_*$, $I$ and $\rho_*$
  are $\sim 30\%$ larger. This results in an increased uncertainty of
  $\sim 30\%$ on the radii and $\sim 20\%$ on the masses.  Hence, we
  conclude that if the orbit eccentricity is $ < 0.038$ the system
  parameters would be within $\sim 1.3 \sigma$ of the values given in
  Tables 2 and 3.

  \section{Discussion and Conclusion}
We have presented two high quality transit light curves of WASP-21b
taken with RISE. Together with the previous RISE partial transit,
these were fitted with an MCMC procedure to update the parameters of
the system. We have been conservative in our error estimates by
scaling the $\chi^2$ and by including time correlated noise in our
analysis.

The derived stellar density $ \rho_* = 0.65 \pm 0.05 \rho_{\odot} $
and the estimated age for the system, $12 \pm 2\,$Gyr, suggest that
WASP-21 is in the process of evolving off the main sequence.
Therefore, the main-sequence mass-radius relation assumed for WASP-21
in the discovery paper was invalid which led to a significant
overestimation of the stellar density, thus affecting the derived
planetary properties. Using the stellar models of
\citet{Demarque2004}, we derived a significantly lower stellar, $M_*=
0.86 \pm 0.03 \Msun$, and planetary mass, $ M_p = 0.27 \pm 0.01
\Mjup$. This lower host star mass somewhat compensates the lower
stellar density which results in a stellar radius which is within
$1\sigma$ of the one presented by \citet{Bouchy2010}.

We obtained a slightly larger planetary radius, $R_p= 1.14 \pm 0.04\,$
\Rjup, for WASP-21b than previously reported.  \citet{Fortney2007}
hydrogen/helium coreless models predict a radius of $\sim 1.06$ \Rsun\
which is consistent within $2\sigma$ with our estimated radius without
the need for any extra heating mechanism. Following
\citet{Laughlin2011} we compute a radius anomaly, $\Re=0.09$, for
WASP-21b. This supports the correlation reported by
\citet{Laughlin2011}, i.e. $\Re =T_{equ}^{1.4}$, where $T_{equ}$ is
the equilibrium temperature of the planet, which is $\sim 1320\,$K for
WASP-21b. \citet{Bouchy2010} argued that the density of WASP-21b
strengthens the correlation between planetary density and host star
metallicity for hot Saturns \citep{Guillot2006}. With the addition of
the latest Saturn-mass planet discoveries (e.g. WASP-39,
\citealt{Faedi2011}; WASP-40, \citealt{Anderson2011}) this correlation
appears weaker. However, if we scale for the equilibrium temperature
with, for example $\Re$, the correlation with metallicity is still
strong (see Figure 6 in \citealt{Faedi2011}).  Moreover, the
correlation also holds for the more massive planets (see Figure 3 in
\citealt{Laughlin2011}).

Exoplanet transit light curves are often affected by systematic noise
that can in some cases dominate the photometric noise. Therefore, it
is important to minimise the sources of systematic noise.  In Section
3.1, we show an example of systematic noise present in our exoplanet
transit observations and suggest that the first step to decrease this
noise is to maintain the star in the same pixel position in the CCD
during the observations. We confirm that defocused observations can
also help decreasing systematic noise, as well decreasing deadtime
losses and hence improving the signal-to-noise
\citep{Southworth2009}. The systematic noise in our observations was
due to the variation of the stellar position across the CCD.

\section{Acknowledgements}
FPK is grateful to AWE Aldermaston for the award of a William Penny
Fellowship. The RISE instrument mounted at the Liverpool Telescope was
designed and built with resources made available from Queen's
University Belfast, Liverpool John Moores University and the
University of Manchester.  The Liverpool Telescope is operated on the
island of La Palma by Liverpool John Moores University in the Spanish
Observatorio del Roque de los Muchachos of the Instituto de
Astrofisica de Canarias with financial support from the UK Science and
Technology Facilities Council. We thank Tom Marsh for the use of the
ULTRACAM pipeline.  SCCB is grateful to Catherine Walsh for
proofreading this paper and to Yilen Gómez Maqueo Chew for useful
comments.

\bibliographystyle{mn2e} \bibliography{susana_mn}

\begin{thebibliography}{}

\bibitem[\protect\citeauthoryear{{Anderson}, {Barros}, {Boisse}, {Bouchy},
  {Collier-Cameron}, {Faedi}, {Hebrard}, {Hellier}, {Lendl}, {Moutou},
  {Pollacco}, {Santerne}, {Smalley}, {Smith}, {Todd} \& {et al.}}{{Anderson}
  et~al.}{2011}]{Anderson2011}
{Anderson} D.~R.,  {Barros} S.~C.~C.,  {Boisse} I.,  {Bouchy} F.,
  {Collier-Cameron} A.,  {Faedi} F.,  {Hebrard} G.,  {Hellier} C.,  {Lendl} M.,
   {Moutou} C.,  {Pollacco} D.,  {Santerne} A.,  {Smalley} B.,  {Smith}
  A.~M.~S.,  {Todd} I.,    {et al.} 2011, ArXiv e-prints

\bibitem[\protect\citeauthoryear{{Bouchy}, {Hebb}, {Skillen}, {Collier
  Cameron}, {Smalley}, {Udry}, {Anderson}, {Boisse}, {Enoch}, {Haswell},
  {H{\'e}brard}, {Hellier}, {Joshi}, {Kane}, {Maxted}, {Mayor}, {Moutou},
  {Pepe} \& {et al.}}{{Bouchy} et~al.}{2010}]{Bouchy2010}
{Bouchy} F.,  {Hebb} L.,  {Skillen} I.,  {Collier Cameron} A.,  {Smalley} B.,
  {Udry} S.,  {Anderson} D.~R.,  {Boisse} I.,  {Enoch} B.,  {Haswell} C.~A.,
  {H{\'e}brard} G.,  {Hellier} C.,  {Joshi} Y.,  {Kane} S.~R.,  {Maxted} P.~F.,
   {Mayor} M.,  {Moutou} C.,  {Pepe} F.,    {et al.} 2010, \aap, 519, A98+

\bibitem[\protect\citeauthoryear{{Carter} \& {Winn}}{{Carter} \&
  {Winn}}{2009}]{carter2009}
{Carter} J.~A.,  {Winn} J.~N.,  2009, \apj, 704, 51

\bibitem[\protect\citeauthoryear{{Collier Cameron}, {Wilson}, {West}, {Hebb},
  {Wang}, {Aigrain}, {Bouchy}, {Christian}, {Clarkson}, {Enoch}, {Esposito},
  {Guenther}, {Haswell}, {H{\'e}brard} \& {et al.}}{{Collier Cameron}
  et~al.}{2007}]{Cameron2007}
{Collier Cameron} A.,  {Wilson} D.~M.,  {West} R.~G.,  {Hebb} L.,  {Wang} X.,
  {Aigrain} S.,  {Bouchy} F.,  {Christian} D.~J.,  {Clarkson} W.~I.,  {Enoch}
  B.,  {Esposito} M.,  {Guenther} E.,  {Haswell} C.~A.,  {H{\'e}brard} G.,
  {et al.} 2007, \mnras, 380, 1230

\bibitem[\protect\citeauthoryear{{Demarque}, {Woo}, {Kim} \& {Yi}}{{Demarque}
  et~al.}{2004}]{Demarque2004}
{Demarque} P.,  {Woo} J.,  {Kim} Y.,    {Yi} S.~K.,  2004, \apjs, 155, 667

\bibitem[\protect\citeauthoryear{{Dhillon}, {Marsh}, {Stevenson}, {Atkinson},
  {Kerry}, {Peacocke}, {Vick}, {Beard}, {Ives}, {Lunney}, {McLay}, {Tierney},
  {Kelly}, {Littlefair}, {Nicholson}, {Pashley}, {Harlaftis} \&
  {O'Brien}}{{Dhillon} et~al.}{2007}]{Ultracam}
{Dhillon} V.~S.,  {Marsh} T.~R.,  {Stevenson} M.~J.,  {Atkinson} D.~C.,
  {Kerry} P.,  {Peacocke} P.~T.,  {Vick} A.~J.~A.,  {Beard} S.~M.,  {Ives}
  D.~J.,  {Lunney} D.~W.,  {McLay} S.~A.,  {Tierney} C.~J.,  {Kelly} J.,
  {Littlefair} S.~P.,  {Nicholson} R.,  {Pashley} R.,  {Harlaftis} E.~T.,
  {O'Brien} K.,  2007, \mnras, 378, 825

\bibitem[\protect\citeauthoryear{{Enoch}, {Collier Cameron}, {Parley} \&
  {Hebb}}{{Enoch} et~al.}{2010}]{Enoch2010}
{Enoch} B.,  {Collier Cameron} A.,  {Parley} N.~R.,    {Hebb} L.,  2010, \aap,
  516, A33+

\bibitem[\protect\citeauthoryear{{Faedi}, {Barros}, {Anderson}, {Brown},
  {Collier Cameron}, {Pollacco}, {Boisse}, {Hebrard}, {Lendl}, {Lister},
  {Smalley}, {Street}, {Triaud} \& {et al.}}{{Faedi} et~al.}{2011}]{Faedi2011}
{Faedi} F.,  {Barros} S.~C.~C.,  {Anderson} D.~R.,  {Brown} D.~J.~A.,  {Collier
  Cameron} A.,  {Pollacco} D.,  {Boisse} I.,  {Hebrard} G.,  {Lendl} M.,
  {Lister} T.~A.,  {Smalley} B.,  {Street} R.~A.,  {Triaud} A.~H.~M.~J.,    {et
  al.} 2011, ArXiv e-prints

\bibitem[\protect\citeauthoryear{{Fortney}, {Marley} \& {Barnes}}{{Fortney}
  et~al.}{2007}]{Fortney2007}
{Fortney} J.~J.,  {Marley} M.~S.,    {Barnes} J.~W.,  2007, \apj, 659, 1661

\bibitem[\protect\citeauthoryear{Gelman \& Rubin}{Gelman \&
  Rubin}{1992}]{Gelman92}
Gelman A.,  Rubin D.,  1992, Statistical Science, 7, 457

\bibitem[\protect\citeauthoryear{{Gibson}, {Pollacco}, {Simpson}, {Joshi},
  {Todd}, {Benn}, {Christian}, {Hrudkov{\'a}}, {Keenan}, {Meaburn}, {Skillen}
  \& {Steele}}{{Gibson} et~al.}{2008}]{Gibson2008}
{Gibson} N.~P.,  {Pollacco} D.,  {Simpson} E.~K.,  {Joshi} Y.~C.,  {Todd} I.,
  {Benn} C.,  {Christian} D.,  {Hrudkov{\'a}} M.,  {Keenan} F.~P.,  {Meaburn}
  J.,  {Skillen} I.,    {Steele} I.~A.,  2008, \aap, 492, 603

\bibitem[\protect\citeauthoryear{{Gillon}, {Smalley}, {Hebb}, {Anderson},
  {Triaud}, {Hellier}, {Maxted}, {Queloz} \& {Wilson}}{{Gillon}
  et~al.}{2009}]{gillon2009}
{Gillon} M.,  {Smalley} B.,  {Hebb} L.,  {Anderson} D.~R.,  {Triaud}
  A.~H.~M.~J.,  {Hellier} C.,  {Maxted} P.~F.~L.,  {Queloz} D.,    {Wilson}
  D.~M.,  2009, \aap, 496, 259

\bibitem[\protect\citeauthoryear{{Guillot}}{{Guillot}}{2005}]{Guillot2005}
{Guillot} T.,  2005, Annual Review of Earth and Planetary Sciences, 33, 493

\bibitem[\protect\citeauthoryear{{Guillot}, {Santos}, {Pont}, {Iro}, {Melo} \&
  {Ribas}}{{Guillot} et~al.}{2006}]{Guillot2006}
{Guillot} T.,  {Santos} N.~C.,  {Pont} F.,  {Iro} N.,  {Melo} C.,    {Ribas}
  I.,  2006, \aap, 453, L21

\bibitem[\protect\citeauthoryear{{Howarth}}{{Howarth}}{2011}]{Howarth2010}
{Howarth} I.~D.,  2011, \mnras, 413, 1515

\bibitem[\protect\citeauthoryear{{Joshi}, {Pollacco}, {Cameron}, {Skillen},
  {Simpson}, {Steele}, {Street}, {Stempels}, {Christian}, {Hebb}, {Bouchy},
  {Gibson}, {H{\'e}brard}, {Keenan}, {Loeillet} \& {et al.}}{{Joshi}
  et~al.}{2009}]{Joshi2009}
{Joshi} Y.~C.,  {Pollacco} D.,  {Cameron} A.~C.,  {Skillen} I.,  {Simpson} E.,
  {Steele} I.,  {Street} R.~A.,  {Stempels} H.~C.,  {Christian} D.~J.,  {Hebb}
  L.,  {Bouchy} F.,  {Gibson} N.~P.,  {H{\'e}brard} G.,  {Keenan} F.~P.,
  {Loeillet} B.,    {et al.} 2009, \mnras, 392, 1532

\bibitem[\protect\citeauthoryear{{Laughlin}, {Crismani} \& {Adams}}{{Laughlin}
  et~al.}{2011}]{Laughlin2011}
{Laughlin} G.,  {Crismani} M.,    {Adams} F.~C.,  2011, \apjl, 729, L7+

\bibitem[\protect\citeauthoryear{{Lucy} \& {Sweeney}}{{Lucy} \&
  {Sweeney}}{1971}]{Lucy1971}
{Lucy} L.~B.,  {Sweeney} M.~A.,  1971, \aj, 76, 544

\bibitem[\protect\citeauthoryear{{Mandel} \& {Agol}}{{Mandel} \&
  {Agol}}{2002}]{Mandel2002}
{Mandel} K.,  {Agol} E.,  2002, \apjl, 580, L171

\bibitem[\protect\citeauthoryear{{Pollacco}, {Skillen}, {Cameron}, {Christian},
  {Hellier}, {Irwin}, {Lister}, {Street}, {West}, {Anderson}, {Clarkson},
  {Deeg}, {Enoch}, {Evans}, {Fitzsimmons}, {Haswell}, {Hodgkin} \& {et
  al.}}{{Pollacco} et~al.}{2006}]{Pollacco2006}
{Pollacco} D.~L.,  {Skillen} I.,  {Cameron} A.~C.,  {Christian} D.~J.,
  {Hellier} C.,  {Irwin} J.,  {Lister} T.~A.,  {Street} R.~A.,  {West} R.~G.,
  {Anderson} D.,  {Clarkson} W.~I.,  {Deeg} H.,  {Enoch} B.,  {Evans} A.,
  {Fitzsimmons} A.,  {Haswell} C.~A.,  {Hodgkin} S.,    {et al.} 2006, \pasp,
  118, 1407

\bibitem[\protect\citeauthoryear{{Simpson}, {Faedi}, {Barros}, {Brown},
  {Collier Cameron}, {Hebb}, {Pollacco}, {Smalley}, {Todd}, {Butters},
  {H{\'e}brard}, {McCormac}, {Miller}, {Santerne}, {Street}, {Skillen},
  {Triaud} \& {et al.}}{{Simpson} et~al.}{2011}]{Simpson2011}
{Simpson} E.~K.,  {Faedi} F.,  {Barros} S.~C.~C.,  {Brown} D.~J.~A.,  {Collier
  Cameron} A.,  {Hebb} L.,  {Pollacco} D.,  {Smalley} B.,  {Todd} I.,
  {Butters} O.~W.,  {H{\'e}brard} G.,  {McCormac} J.,  {Miller} G.~R.~M.,
  {Santerne} A.,  {Street} R.~A.,  {Skillen} I.,  {Triaud} A.~H.~M.~J.,    {et
  al.} 2011, \aj, 141, 8

\bibitem[\protect\citeauthoryear{{Southworth}}{{Southworth}}{2008}]{Southworth2008}
{Southworth} J.,  2008, \mnras, 386, 1644

\bibitem[\protect\citeauthoryear{{Southworth}, {Hinse}, {J{\o}rgensen},
  {Dominik}, {Ricci}, {Burgdorf}, {Hornstrup}, {Wheatley}, {Anguita}, {Bozza}
  \& {et al.}}{{Southworth} et~al.}{2009}]{Southworth2009}
{Southworth} J.,  {Hinse} T.~C.,  {J{\o}rgensen} U.~G.,  {Dominik} M.,  {Ricci}
  D.,  {Burgdorf} M.~J.,  {Hornstrup} A.,  {Wheatley} P.~J.,  {Anguita} T.,
  {Bozza} V.,    {et al.} 2009, \mnras, 396, 1023

\bibitem[\protect\citeauthoryear{{Sozzetti}, {Torres}, {Charbonneau}, {Latham},
  {Holman}, {Winn}, {Laird} \& {O'Donovan}}{{Sozzetti}
  et~al.}{2007}]{Sozzetti2007}
{Sozzetti} A.,  {Torres} G.,  {Charbonneau} D.,  {Latham} D.~W.,  {Holman}
  M.~J.,  {Winn} J.~N.,  {Laird} J.~B.,    {O'Donovan} F.~T.,  2007, \apj, 664,
  1190

\bibitem[\protect\citeauthoryear{{Steele}, {Bates}, {Gibson}, {Keenan},
  {Meaburn}, {Mottram}, {Pollacco} \& {Todd}}{{Steele} et~al.}{2008}]{rise2008}
{Steele} I.~A.,  {Bates} S.~D.,  {Gibson} N.,  {Keenan} F.,  {Meaburn} J.,
  {Mottram} C.~J.,  {Pollacco} D.,    {Todd} I.,  2008, in Society of
  Photo-Optical Instrumentation Engineers (SPIE) Conference Series Vol.~7014 of
  Society of Photo-Optical Instrumentation Engineers (SPIE) Conference Series,
  {RISE: a fast-readout imager for exoplanet transit timing}

\bibitem[\protect\citeauthoryear{{Tegmark}, {Strauss}, {Blanton}, {Abazajian},
  {Dodelson}, {Sandvik}, {Wang}, {Weinberg}, {Zehavi}, {Bahcall}, {Hoyle},
  {Schlegel}, {Scoccimarro}, {Vogeley} \& {et al.}}{{Tegmark}
  et~al.}{2004}]{Tegmark2004}
{Tegmark} M.,  {Strauss} M.~A.,  {Blanton} M.~R.,  {Abazajian} K.,  {Dodelson}
  S.,  {Sandvik} H.,  {Wang} X.,  {Weinberg} D.~H.,  {Zehavi} I.,  {Bahcall}
  N.~A.,  {Hoyle} F.,  {Schlegel} D.,  {Scoccimarro} R.,  {Vogeley} M.~S.,
  {et al.} 2004, \prd, 69, 103501

\bibitem[\protect\citeauthoryear{{Torres}, {Andersen} \&
  {Gim{\'e}nez}}{{Torres} et~al.}{2010}]{Torres2010}
{Torres} G.,  {Andersen} J.,    {Gim{\'e}nez} A.,  2010, \aapr, 18, 67

\end{thebibliography}

\label{lastpage}

\end{document}